\pdfoutput=1

\documentclass[11pt,twoside,a4paper,cmspaper,final,collab]{cms-tdr}
\def\svnVersion{422517P}\def\cmsCernNoTag{CERN-EP-2017-002}\def\cmsCernDate{\today}\def\cmsMessage{Published in Physical Review Letters as \href{http://dx.doi.org/10.1103/PhysRevLett.119.082301}{\doi{10.1103/PhysRevLett.119.082301.}}}

\begin{document}\cmsNoteHeader{HIN-15-013}

\hyphenation{had-ron-i-za-tion}
\hyphenation{cal-or-i-me-ter}
\hyphenation{de-vices}
\RCS$HeadURL: svn+ssh://svn.cern.ch/reps/tdr2/papers/HIN-15-013/trunk/HIN-15-013.tex $
\RCS$Id: HIN-15-013.tex 414378 2017-07-03 18:53:04Z alverson $
\newlength\cmsFigWidth
\ifthenelse{\boolean{cms@external}}{\setlength\cmsFigWidth{0.85\columnwidth}}{\setlength\cmsFigWidth{0.4\textwidth}}
\ifthenelse{\boolean{cms@external}}{\providecommand{\cmsLeft}{top\xspace}}{\providecommand{\cmsLeft}{left\xspace}}
\ifthenelse{\boolean{cms@external}}{\providecommand{\cmsRight}{bottom\xspace}}{\providecommand{\cmsRight}{right\xspace}}
\ifthenelse{\boolean{cms@external}} {\providecommand{\suppMaterial}{the supplemental material[URL will be inserted by publisher]}}{\providecommand{\suppMaterial}{Appendix~\ref{app:suppMat}}}

\cmsNoteHeader{HIN-15-013}

\newcommand {\PbPb}  {\ensuremath{\mathrm{PbPb}}\xspace}
\newcommand {\ptz}    {\ensuremath{p_{\mathrm{T}}^{\Z}}\xspace}
\newcommand {\yz}    {\ensuremath{y^{\Z}}\xspace}
\newcommand {\phiz}    {\ensuremath{\phi^{\Z}}\xspace}
\newcommand {\ptj}    {\ensuremath{p_{\mathrm{T}}^\text{jet}}}\xspace
\newcommand {\etaj}    {\ensuremath{\eta^\text{jet}}\xspace}
\newcommand {\phij}    {\ensuremath{\phi^\text{jet}}\xspace}
\newcommand {\ptGen}    {\ensuremath{\pt^\text{gen}}\xspace}
\newcommand {\ptReco}    {\ensuremath{\pt^\text{reco}}\xspace}
\newcommand {\phiGen}    {\ensuremath{\phi^\text{gen}}\xspace}
\newcommand {\phiReco}    {\ensuremath{\phi^\text{reco}}\xspace}
\newcommand {\eScale}    {\ensuremath{\langle \ptReco / \ptGen \rangle}}
\providecommand{\xjzMean}{\ensuremath{\langle x_{\cmsSymbolFace{jZ}} \rangle}\xspace}
\providecommand{\rjz}{\ensuremath{R_{\cmsSymbolFace{jZ}}}\xspace}
\providecommand{\xjz}{\ensuremath{x_{\cmsSymbolFace{jZ}}}\xspace}
\providecommand{\dphijz}{\ensuremath{\Delta\phi_{\cmsSymbolFace{jZ}}}\xspace}
\providecommand{\rootsNN}{\ensuremath{\sqrt{\smash[b]{s_{_\mathrm{NN}}}}}\xspace}
\providecommand{\HYDJET}{\textsc{hydjet}\xspace}
\providecommand{\MGATNLO}{\textsc{MadGraph}5\_a\textsc{mc@nlo}\xspace}

\title{\texorpdfstring{Study of jet quenching with \Z{}+jet correlations in \PbPb and $\Pp\Pp$ collisions at $\rootsNN = 5.02\TeV$}{Study of jet quenching with Z+jet correlations in PbPb and pp collisions at sqrt(s[NN]) = 5.02 TeV}}

\date{\today}

\abstract{
The production of jets in association with \Z bosons, reconstructed via the $\PGmp\PGmm$ and $\Pep\Pem$ decay channels, is studied in $\Pp\Pp$ and, for the first time, in \PbPb collisions. Both data samples were collected by the CMS experiment at the LHC, at a nucleon-nucleon center-of-mass energy of 5.02\TeV. The \PbPb collisions were analyzed in the 0--30\% centrality range. The back-to-back azimuthal alignment was studied in both $\Pp\Pp$ and \PbPb collisions for \Z bosons with transverse momentum
$\ptz > 60\GeVc$ and a recoiling jet with $\ptj > 30\GeVc$. The \pt imbalance, $\xjz=\ptj/\ptz$, as well as the average number of jet partners per \Z, \rjz, were studied in intervals of \ptz. The \rjz is found to be smaller in \PbPb than in $\Pp\Pp$ collisions, which suggests that in \PbPb collisions a larger fraction of partons associated with the \Z bosons fall below the 30\GeVc \ptj threshold because they lose energy.}

\hypersetup{%
pdfauthor={CMS Collaboration},%
pdftitle={Study of jet quenching with Z+jet correlations in PbPb and pp collisions at sqrt(s[NN]) = 5.02 TeV},%
pdfsubject={CMS},%
pdfkeywords={CMS, physics, heavy ion, jet quenching}}

\maketitle
The correlated production of vector bosons and jets in hard parton scatterings occurring
in ultrarelativistic heavy ion collisions provides an ideal probe
of the quark-gluon plasma (QGP), a deconfined state of quarks and gluons~\cite{Shuryak:1978ij, Shuryak:2004cy}. 
Final-state jets are created by the fragmentation of outgoing partons that interact strongly with the produced medium and lose energy~\cite{Appel:1985dq, Blaizot:1986ma, Gyulassy:1990ye, Wang:1991xy, Baier:1996sk, Zakharov:1997uu,CasalderreySolana:2007zz,d'Enterria:2009am,Wiedemann:2009sh}, a phenomenon (``jet quenching") observed at RHIC~\cite{PhysRevLett.97.162301,PhysRevC.77.011901} and the LHC~\cite{Aad:2010bu, Chatrchyan:2011sx, Chatrchyan2012176,Chatrchyan:2012gt,Adam20151}. The transverse momentum (\pt) of the jet is highly correlated (through momentum conservation) with that of the associated \Z boson, which is not affected by the medium~\cite{Kartvelishvili:1995fr, Neufeld:2012df, Neufeld:2010fj} and reflects the initial energy of the parton. 
The lost energy can be related, via theoretical models, to the thermodynamical and transport properties of the medium~\cite{CasalderreySolana:2007zz,d'Enterria:2009am,Wiedemann:2009sh,CasalderreySolana:2012ef,PhysRevC.90.014909,Casalderrey-Solana:2014bpa}. 
At LHC energies, \Z{}+jet production is dominated by quark jets for $\ptj \gtrsim 30$\GeVc~\cite{Neufeld:2010fj}, the primary subprocess being $\PQq( \PAQq ) + \Pg \to \Z +\PQq( \PAQq)$~\cite{Kartvelishvili:1995fr},
hence providing information on the parton flavor (quark or gluon) and kinematics, and allowing detailed studies of the energy loss with a well-defined production process. 
The \Z{}-jet correlations
are particularly well suited to perform tomographic studies of the QGP, given the minimal contributions from background channels~\cite{Neufeld:2012df, Neufeld:2012df, Casalderrey-Solana2016, KunnawalkamElayavalli:2016ttl, Kang:2017xnc}.
Correlations of jets with isolated photons are accessible at higher rates
and carry similar information on parton energy loss~\cite{Wang:1996yh, Wang:1996pe, Casalderrey-Solana2016, KunnawalkamElayavalli:2016ttl, Kang:2017xnc},
but suffer from an irreducible background of photons from jet
fragmentation~\cite{Chatrchyan:2012vq, Chatrchyan:2012gt}, as well as larger uncertainties arising
from the experimental selection of photon candidates. 

This Letter describes the identification of \Z{}+jet pairs in $\Pp\Pp$ and \PbPb collisions, and the first characterization of parton energy loss through angular and \pt correlations
between the jet and the \Z, reconstructed in dimuon or dielectron decays. The back-to-back azimuthal alignment of the \Z and jets is studied through the difference $\dphijz = \abs{\phij - \phiz}$. The \Z{}+jet momentum imbalance is studied using the $\xjz =\ptj / \ptz$ ratio and the \ptz dependence of its mean value, \xjzMean. The average number of jet partners per \Z boson, \rjz, is also reported. The analysis exploits \PbPb and $\Pp\Pp$ data samples collected by CMS at a nucleon-nucleon center-of-mass energy of 5.02\TeV, corresponding to integrated luminosities of 404\mubinv and 27.4\pbinv, respectively.

The central feature of CMS is a superconducting solenoid of 6\unit{m} internal diameter, providing a magnetic field of 3.8\unit{T}. Within the solenoid volume are a silicon pixel and strip tracker, a lead tungstate crystal electromagnetic calorimeter (ECAL), and a brass and scintillator hadron calorimeter (HCAL), each composed of a barrel and two endcap sections. Forward hadron calorimeters extend the pseudorapidity ($\eta$) coverage and are used for \PbPb event selection. Muons are measured in gas-ionization detectors located outside the solenoid. A more detailed description of the CMS detector, together with a definition of the coordinate system used and the relevant kinematic variables, can be found in Ref.~\cite{Chatrchyan:2008zzk}.

The event samples are selected online with dedicated lepton triggers, and
cleaned offline to remove noncollision events, such as beam-gas interactions
or cosmic-ray muons~\cite{Khachatryan:2016odn}. In addition, events are required to have at least one reconstructed primary interaction vertex. 
The $\Z \to \Pep\Pem$ events are triggered if two ECAL clusters~\cite{Khachatryan:2016bia} have transverse energy greater than 15\GeV and $\abs{\eta}< 2.5$,
while the $\Z \to \PGmp\PGmm$ triggers require one muon of $\pt > 15\GeVc$ or two muons of $\pt > 10\GeVc$.

For the analysis of \PbPb collisions, the ``centrality'' (overlap of the two colliding nuclei) is determined by the sum of the total energy deposited in both forward hadron calorimeters~\cite{Chatrchyan:2011sx}. The results refer to the 30\% most central collisions, to focus on the region of highest physics interest. After all the other analysis selections, 78\% of the \Z boson events fall in this centrality range.

The \PYTHIA~8.212~\cite{pythia8} Monte Carlo (MC) event generator, with tune CUETP8M1~\cite{Khachatryan:2015pea},
is used to simulate \Z{}+jet signal events, with $\ptz > 30\GeVc$ and rapidity $\abs{\yz}<2.5$. A sample with a \Z boson without any kinematic selection was produced using a next-to-leading order (NLO) generator, \MGATNLO~\cite{Alwall:2014hca}. In the \PbPb case, a {\PYTHIA{}+\HYDJET} sample is created by embedding \PYTHIA signal events in heavy ion events generated with {\HYDJET}~1.9~\cite{Lokhtin:2005px} and tune HydroQJets. The generated events are propagated through the CMS apparatus using the \GEANTfour~\cite{geant4} package. No unfolding is performed for the results presented. The recipe for applying a smearing of the jet \pt resolution is provided in~\suppMaterial.

Electrons are identified as ECAL superclusters~\cite{1748-0221-10-06-P06005} matched in position and energy to tracks reconstructed in the tracker. They must have $\pt > 20\GeVc$, above the trigger threshold, and each supercluster must 
be within the acceptance of the tracker, $\abs{\eta}< 2.5$.  Electron candidates in the transition region between the barrel and endcap subdetectors ($1.44 < \abs{\eta}< 1.57$) are excluded. In $\Pp\Pp$ collisions, the electrons are
selected via standard identification criteria~\cite{1748-0221-10-06-P06005}. 
A narrow transverse shape of showers in the ECAL and a low 
HCAL over ECAL energy ratio are required to reject misidentified electrons. Additional tracking information is used to distinguish electrons from charged hadrons~\cite{1748-0221-10-06-P06005}. For \PbPb collisions, the identification criteria have been optimized to compensate for the higher background levels in the calorimeters. With these selections, the $\Pp\Pp$ and \PbPb electron reconstruction purities (efficiencies) are identical within 1\% (10\%).

Muons are selected by requiring segments in at least two muon detector planes and a good-quality fit when connecting them to tracker segments. This suppresses hadronic punch-through and muons from in-flight decays of hadrons. A minimum number of hits in the pixel and strip 
layers is required, and the reconstructed muon
tracks must point to the primary vertex in the transverse and longitudinal
directions~\cite{Chatrchyan:2012xi}. The same selections are applied for both
$\Pp\Pp$ and \PbPb data. In order to suppress the background continuum under the \Z peak, mostly originating from uncorrelated simultaneous decays of heavy flavour mesons, the muons are required to have $\pt > 10\GeVc$. In addition, the muon tracks must fall in the acceptance of the muon detectors, $\abs{\eta^\mu}<2.4$.

Jet reconstruction uses the
anti-\kt algorithm implemented in {\FASTJET}~\cite{Cacciari:2011ma}, following the procedure of Ref.~\cite{Chatrchyan2012176}. A small distance parameter, $R= 0.3$, minimizes the effects of fluctuations in the underlying event (UE), dominantly formed by
soft processes in heavy ion collisions. The UE energy subtraction~\cite{Kodolova:2007hd} is performed for \PbPb as described in Refs.~\cite{Chatrchyan:2011sx,Chatrchyan:2012gt,Chatrchyan2012176}.
Closure tests, done on MC samples without medium induced jet energy loss, show no over subtraction of the UE in the \PbPb sample. No subtraction is applied in the $\Pp\Pp$ sample, where the UE contribution is negligible.
The jet energy is calibrated applying $\etaj$- and \ptj-dependent correction factors derived with the \PYTHIA signal sample~\cite{1748-0221-6-11-P11002}.
Then, dijet and photon+jet balance techniques~\cite{Khachatryan:2015lha} are used
to correct for the residual detector
response differences between
measured and simulated samples.
In addition, a centrality-dependent correction obtained from simulation studies
is applied to remove the residual effects from the UE in \PbPb collisions.
The UE from \PbPb data and MC samples are compared using the \pt density~\cite{1748-0221-6-11-P11002,Cacciari2011,Cacciari2008119}, defined as the median of the ratio of the jet transverse momentum to the jet area, for all jets in the event.
Given the coarse centrality range used in the analysis, the difference between the measured and simulated \PbPb events
has a negligible effect on jet reconstruction.

\begin{figure}[th]\centering
\includegraphics[width=0.43\textwidth]{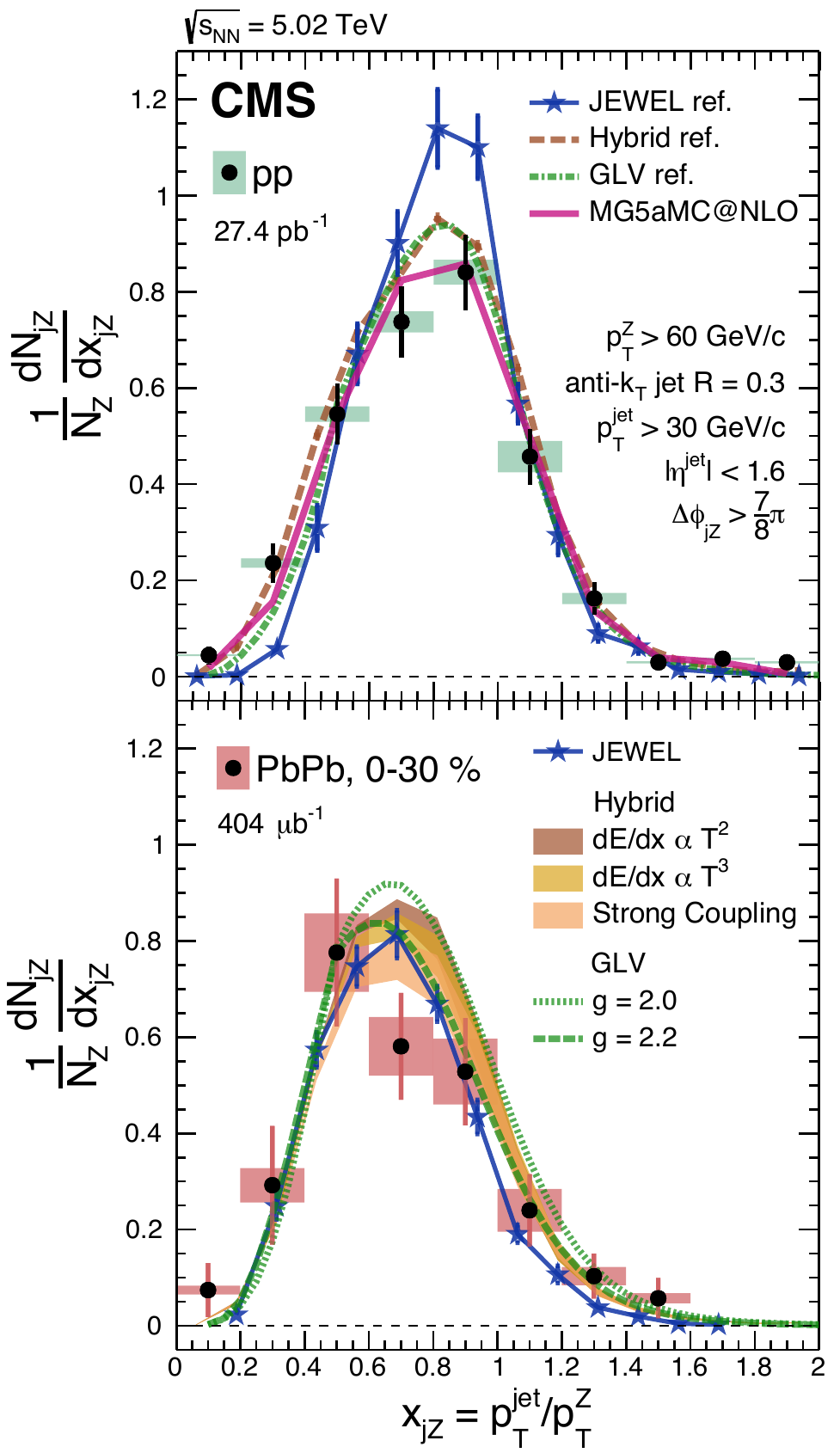}
\caption{\label{fig:xjz_theory}
Comparison of the measured $\Pp\Pp$ (top) and \PbPb (bottom) \xjz distributions with several theoretical models, smeared by the respective jet energy resolution:
JEWEL~\cite{KunnawalkamElayavalli:2016ttl}, Hybrid~\cite{Casalderrey-Solana2016}, and GLV~\cite{Kang:2017xnc}.
The JEWEL error bars represent statistical uncertainties while the widths of the Hybrid bands represent systematic variations.
A \MGATNLO calculation~\cite{Alwall:2014hca} is also shown.}
\end{figure}

Except in Fig.~\ref{fig:xjz_theory}, the resolutions of the measured jet energy and azimuthal angle in the $\Pp\Pp$ samples are smeared to match those of the \PbPb sample. 
The jet energy resolution can be quantified using the Gaussian standard deviation $\sigma$ of the
$\ptReco/\ptGen$ ratio, where $\ptReco$ is the
UE-subtracted, detector-level jet \pt and $\ptGen$
is the generator-level jet \pt  without any contributions from the UE in \PbPb.
It is determined using {\PYTHIA{}+\HYDJET} (for \PbPb) and \PYTHIA (for $\Pp\Pp$) samples and parametrized as a function of $\ptGen$ using the expression $\sigma\left(\ptGen\right) = C \oplus (S/\sqrt{\ptGen}) \oplus (N/{\ptGen})$, where $\oplus$ stands for the sum in quadrature and the parameters $C$, $S$, and $N$ are determined from
simulation studies. The same parametrization is used to determine the jet azimuthal angle resolution, quantified by the Gaussian standard deviation $\sigma_{\phi}$ of the $\abs{\phiReco-\phiGen}$ difference.

\begin{figure}[th]\centering
    \includegraphics[width=0.42\textwidth]{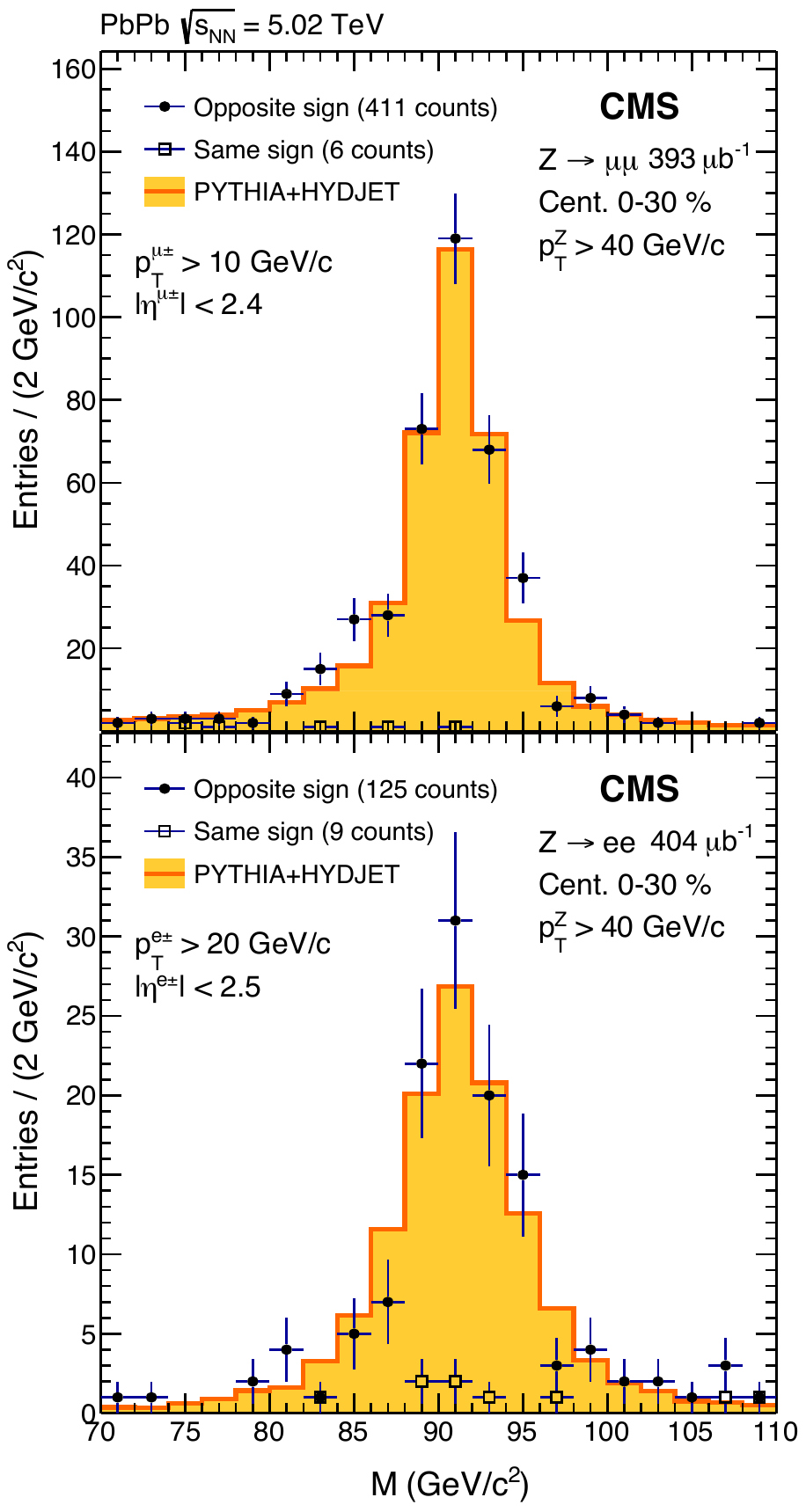}
    \caption{Invariant mass distributions of the selected dimuons (top) and dielectrons (bottom),
    compared to {\PYTHIA{}+\HYDJET} $\Z(\ell\ell)$+jet events. The MC histogram is normalized to the number of events in the data.}
    \label{fig:massfits}
\end{figure}

The \Z candidates are defined as opposite-charge electron or muon pairs, with a reconstructed invariant mass ($M^{\ell\ell}$) in the interval 70--$110\GeVc^2$ and $\pt>40\GeVc$.
The invariant mass distributions of all the dileptons used in the \PbPb analysis are shown in Fig.~\ref{fig:massfits}. Each \Z candidate is paired with all jets in the same event that pass the $\ptj > 30\GeVc$ and $\abs{\etaj} < 1.6$ selection. Simulation studies show that the jet selection efficiency and the energy resolution are well understood for this kinematic range. Additional energy corrections are applied to the jet \pt, to account for residual performance degradations observed in simulation studies. 
Jets reconstructed within $\Delta R < 0.4$ from a lepton are rejected,
to eliminate jet energy contamination by leptons from \Z decays.

For the analysis of \PbPb collisions, the background contribution from jets not produced in the same parton-parton interaction as the \Z boson needs to be considered. This contribution arises from misidentified jets reconstructed from regional energy fluctuations in the high-multiplicity heavy ion UE, or from additional initial hard interactions not related to the primary \Z{}+jet production. The background jet contributions are estimated constructing a mixed-event jet background by correlating the \Z boson from each candidate \Z{}+jet event with jets reconstructed in subsets of 40 minimum bias events. All events must pass the offline event selection and have the same centrality and interaction vertex as the \Z{}+jet candidate event. The resulting background jet spectrum is subtracted from the raw jet spectrum, eliminating coincidental \Z{}+jet pairs and ensuring that the final \Z{}+jet observables reflect the correlations of \Z bosons and associated jets.

The systematic uncertainties related to \Z boson reconstruction are sizable (negligible) in the dielectron (dimuon) channel. 
Comparing the measured and simulated dielectron invariant mass peaks shows that the average deviation between electron \ptReco and \ptGen is 0.5\%. A systematic uncertainty is evaluated by shifting the electron \pt by
$\pm$0.5\%, resulting in changes of \xjzMean and \rjz for \PbPb ($\Pp\Pp$) by 0.5\% (0.3\%) and 3\% (0.8\%), respectively.
The simulated \Z dielectrons reconstructed in central \PbPb collisions have a \pt resolution of 5\% for $\ptz > 40\GeVc$. In \PbPb simulated events, \ptz is smeared by 5\%, resulting in variations of \xjzMean and \rjz by 1.5\% and 0.8\%, respectively. When combining the two lepton results, a weighting is applied to the electron sample, to compensate for the different centrality dependencies of the \Z boson reconstruction in the electron and muon channels. The difference between the corrected and uncorrected \xjzMean and \rjz values, 0.3\% and 5.8\% respectively, is taken as systematic uncertainty.

Simulation studies show that the jet energy scale, $\eScale$, can deviate from unity by up to 2\%. Additional deviations can arise from differences between the fragmentation pattern of jets in
measured and simulated events.
To evaluate the corresponding systematic uncertainty,
the jet energy scale is shifted for \PbPb ($\Pp\Pp$)  upward by 6\% (2\%) and downward by 4\% (2\%). 
The higher upward variation reflects the relatively high energy scale of quark jets, which contribute more to the \Z{}+jet events than the gluon jets. The relative change in \xjzMean and \rjz for \PbPb ($\Pp\Pp$) is 5.4\% (2.4\%) and 4.6\% (2.4\%), respectively.
Finally, differences between the measured and simulated samples
suggest that the jet energy resolution is up to 15\% worse in data.
The related systematic uncertainty is evaluated smearing $\ptj$ by 15\% in the \PbPb MC. The $\Pp\Pp$ data are smeared to simulate the poor resolution due to the UE fluctuations in \PbPb data. 
The smearing is performed with the relative resolution, $\sigma_{\text{rel}} = \sqrt{\smash[b]{\sigma^{2}_{\PbPb} - \sigma^{2}_{\Pp\Pp}}}$, where $\sigma_{\PbPb}$\ and $\sigma_{\Pp\Pp}$ correspond to the parametrizations described above. 
A systematic uncertainty is assigned by varying the relative resolution by $\pm$\,15\%. 
The \PbPb ($\Pp\Pp$) relative change in \xjzMean and \rjz due to jet energy resolution is 2.5\% and 3.7\% (0.5\% and 0.7\%), respectively.
The jet angular resolution correction implies an additional uncertainty on the $\Pp\Pp$ sample,
of 0.1\% for \xjzMean and 0.2\% for \rjz.

The total systematic uncertainties for \PbPb ($\Pp\Pp$) amount to 6.2\% (2.5\%) and 8.9\% (2.6\%) for the
\xjzMean and \rjz results, respectively, of which 5.7\% and 8.0\% are uncorrelated between the $\Pp\Pp$ and \PbPb results;
the uncorrelated uncertainties do not reflect possible differences between \ptReco and \ptGen.

\begin{figure}[t]\centering
\includegraphics[width=0.42\textwidth]{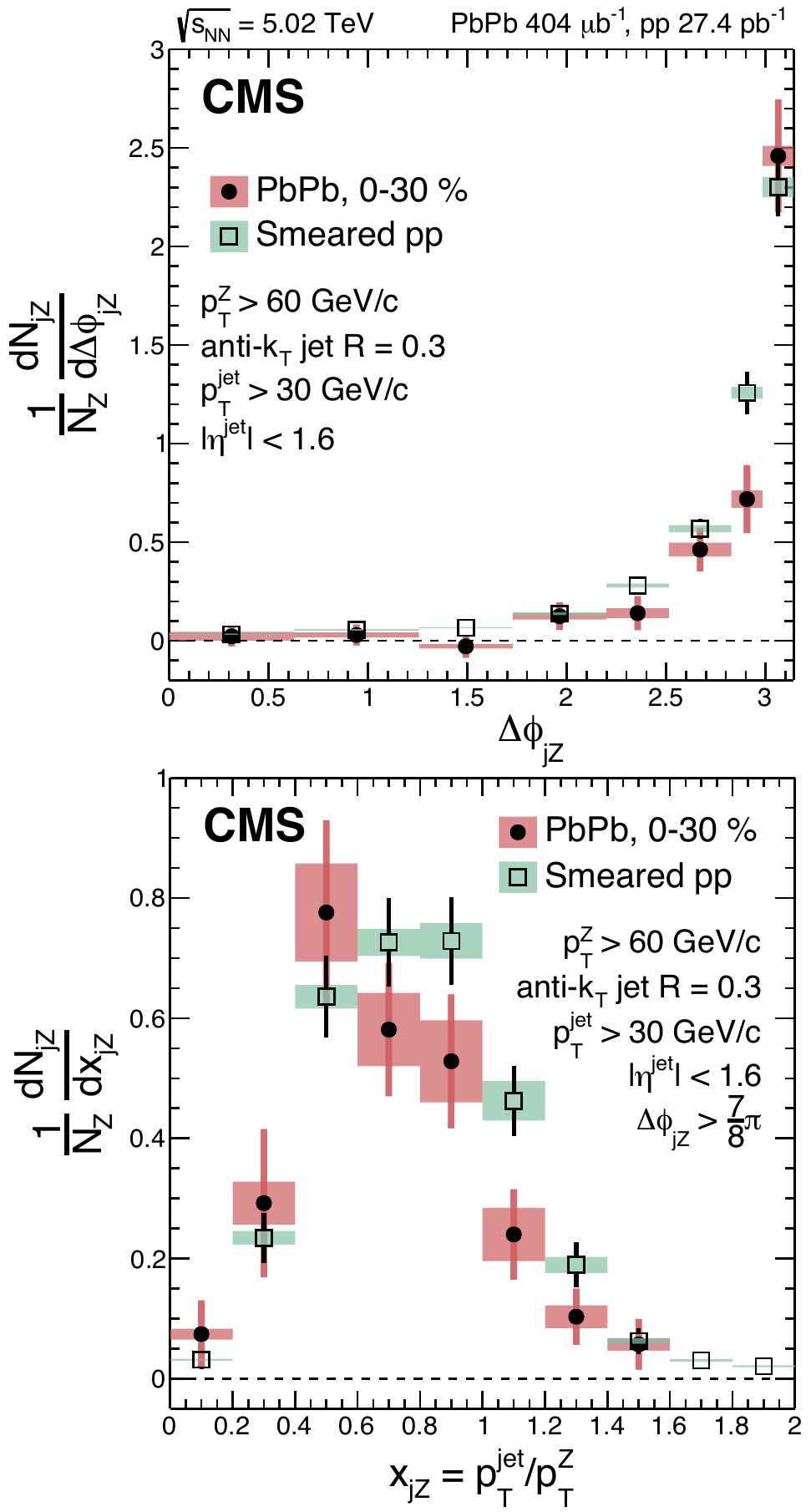}
\caption{\label{fig:dphi_xjz}  Distributions of the azimuthal angle difference \dphijz\ between the \Z boson and the jet (top), and of the transverse momentum ratio \xjz between the jet and the \Z boson with $\dphijz > 7 \pi/8$ (bottom). The distributions are normalized by the number of \Z events, $N_{\Z}$. Vertical lines (bands) indicate statistical (systematic) uncertainties.}
\end{figure}

\begin{figure}[t]\centering
\includegraphics[width=0.41\textwidth]{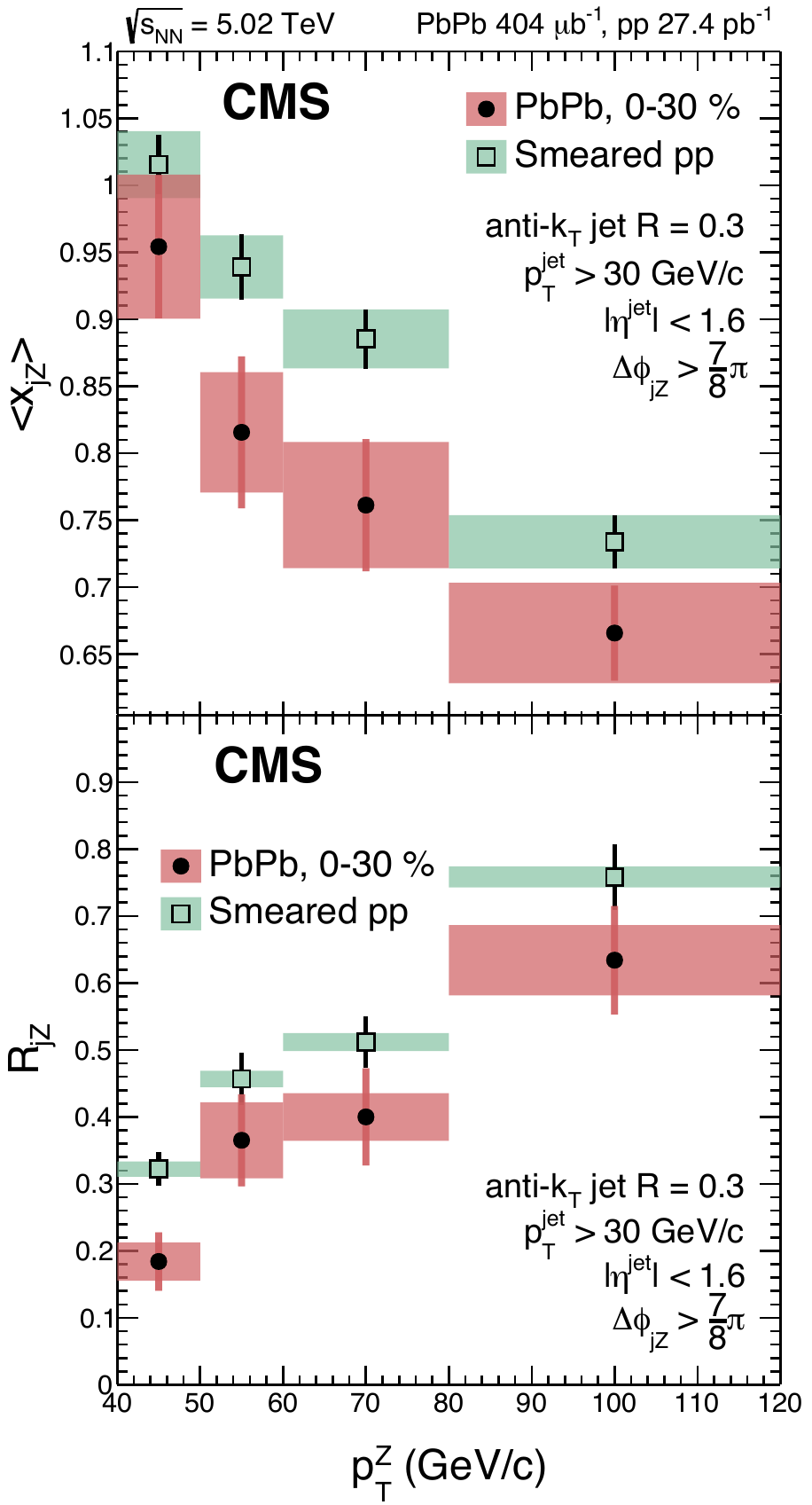}
\caption{\label{fig:xjz_mean_rjz} The mean value of the $\xjz$ distribution (top)
and the average number of jet partners per \Z boson $\rjz$ (bottom), as a function of \ptz.
Vertical lines (bands) indicate statistical (systematic) uncertainties.}
\end{figure}

Figure~\ref{fig:dphi_xjz}-top shows the \dphijz\ distribution of \Z{}+jet
pairs that pass all the selections;
only \Z{}+jet pairs with $\ptz>60$\GeVc were included to reduce the fraction of events where energy loss effects cause the jet partner to fall below the $\ptj>30$\GeVc threshold. There are 678 and 232 events that pass the $\ptz>60$\GeVc selection in $\Pp\Pp$ and in the 30\% most central \PbPb collisions, respectively. To study if the angular distribution of jets with respect to the \Z boson is affected by interactions of the parton with the medium, a Kolmogorov--Smirnov (KS) test was performed using pseudo data generated from identical underlying shapes. This test is useful to quantify shape differences since it is sensitive to adjacent bins fluctuating in the same direction but not to the overall normalization.
No significant difference is seen between the $\Pp\Pp$ and \PbPb \dphijz\ distributions; the probability to obtain a KS value larger than that observed in the data, p-value, is greater than 0.40, even if systematic uncertainties are excluded.

For the \xjz and \rjz results, shown in Figs.~\ref{fig:dphi_xjz} and~\ref{fig:xjz_mean_rjz}, only events with $\dphijz\ > 7\pi/8$ are used,
to select mostly back-to-back \Z{}+jet pairs; it keeps 63\% and 73\% of the $\Pp\Pp$ and \PbPb events, respectively. 
Figure~\ref{fig:dphi_xjz}-bottom shows the \xjz distributions for \PbPb and $\Pp\Pp$ collisions. Jet energy loss is expected to manifest itself both as a shift in the \xjz distribution and an overall decrease in the number of \Z{}+jet pairs as jets fall below the $\ptj$ threshold. Therefore, the KS test was applied to the \xjz distribution and a separate overall normalization $\chi^{2}$ test was applied to the total number of \Z{}+jet pairs per \Z leading to p-values of $p_1 = 0.07$ and $p_2 = 0.01$, respectively. 
The systematic uncertainties and their correlations were included in these calculations. 
The combined $p$-value~\cite{James:2006zz} is $p_1 \, p_2 \, (1 - \ln(p_1 \, p_2)) = 0.0064$
when including \Z{}+jet pairs with $\ptz>40$\GeVc, indicating that the two \xjz distributions are significantly different.

The relative shift between the $\Pp\Pp$ and \PbPb \xjz distributions is studied using their means, $\xjzMean$, 
shown in Fig.~\ref{fig:xjz_mean_rjz}-top as a function of \ptz.
The minimum \pt of the partner jet imposes a lower limit on the value of \xjz. As \ptz increases relative to the $\ptj$ cutoff, the kinematic phase space for lower \xjz opens up, resulting in a shift towards lower \xjz for higher \ptz. For all ranges, $\xjzMean$ is found to be lower in \PbPb collisions than in $\Pp\Pp$ collisions, as expected from energy loss models of partons traversing the medium.
Also \rjz is expected to increase as a
function of \ptz, as the $\ptj > 30$\GeVc threshold restricts the phase space of jets counted for
a given \ptz selection. 
Figure~\ref{fig:xjz_mean_rjz}-bottom shows the dependence of $\rjz$ on \ptz. 
The $\rjz$ values are found to be smaller in \PbPb than in $\Pp\Pp$. 
As their difference is approximately constant as a function of \ptz, a relatively smaller fraction of jets is lost in \PbPb collisions for larger initial (before traversing the medium) parton energies.

Figure~\ref{fig:xjz_theory} compares the \xjz results to several theoretical calculations, using the same kinematic selections as the data. 
The \PbPb results are compared to three models that incorporate the phenomenon of jet quenching: JEWEL~\cite{KunnawalkamElayavalli:2016ttl}, Hybrid~\cite{Casalderrey-Solana2016}, and GLV~\cite{Kang:2017xnc}. 
The \xjz $\Pp\Pp$ measured results are compared to several non-quenching scenarios: 
the $\Pp\Pp$ references used as inputs to the \PbPb models (different tunes of the \textsc{pythia} LO event generator)
and the \MGATNLO generator~\cite{Alwall:2014hca}, which includes matrix elements for \Z plus 0, 1, and 2 jets at NLO, and \Z{}+3 jets at LO. 
The $\Pp\Pp$ calculations were smeared to reflect the detector resolution affecting the $\Pp\Pp$ data. 
The \texttt{JEWEL} model is a dynamical, perturbative framework for jet quenching, which has been extended to simulate boson-jet events~\cite{KunnawalkamElayavalli:2016ttl}. 
This \PbPb \xjz calculation is consistent with the data within the current precision, despite the poor agreement of its baseline with the $\Pp\Pp$ measurement. 
The baseline \textsc{pythia8} tunes used by the Hybrid~\cite{Casalderrey-Solana2016} and GLV~\cite{Kang:2017xnc} models, 
as well as \MGATNLO, describe the $\Pp\Pp$ data reasonably well. 
For \PbPb collisions, the Hybrid model calculation labeled 'Strong Coupling' combines a perturbative description of the weakly coupled physics of jet production and evolution, with a gauge/gravity duality description of the strongly coupled dynamics of the medium, and of the soft exchanges between the jet and the medium. 
Two weak coupling benchmark calculations are also shown, where the energy loss has a quadratic temperature dependence (collisional energy loss), or a cubic dependence (radiative energy loss). 
Given the large experimental and theoretical uncertainties, all three scenarios describe the \PbPb data reasonably well and cannot be distinguished. 
Nevertheless, the Strong Coupling curve appears closest to the data. 
The GLV model~\cite{Kang:2017xnc} generates the energy loss via out-of-cone radiation and collisional energy dissipation. 
Two curves are shown, for different coupling strengths between the jet and the medium, $g = 2.0$ and 2.2, 
reflecting previous analyses of jet quenching measurements at 2.76\TeV~\cite{Vitev:2016pnj,Chien:2015hda};
the $g=2.2$ curve seems favored by the data.

In summary, correlations of $\ptz>40$\GeVc \Z bosons with $\ptj>30$\GeVc jets have been studied in $\Pp\Pp$ and, for the first time, in \PbPb collisions. The data were collected with the CMS experiment
during the 2015 data taking period, at $\rootsNN= 5.02\TeV$. No significant difference was found between the distributions of the
azimuthal angle difference of the \Z and the jet in $\Pp\Pp$ and \PbPb collisions. The \xjz distributions indicate that the \PbPb values tend to be lower than those measured in $\Pp\Pp$ collisions. Correspondingly, the average value of the transverse momentum ratio $\xjzMean$ is smaller in \PbPb than in $\Pp\Pp$ collisions, for all \ptz intervals. The average number of jet partners per \Z, \rjz, is lower in \PbPb than in $\Pp\Pp$ collisions, for all \ptz intervals, which suggests that in \PbPb collisions a larger fraction of partons associated with \Z bosons lose energy and fall below the 30\GeVc \ptj threshold. These measurements provide new input for the determination of jet quenching parameters using a selection of partons with well-defined flavor and initial kinematics.

\begin{acknowledgments}
We congratulate our colleagues in the CERN accelerator departments for the excellent performance of the LHC and thank the technical and administrative staffs at CERN and at other CMS institutes for their contributions to the success of the CMS effort. In addition, we gratefully acknowledge the computing centers and personnel of the Worldwide LHC Computing Grid for delivering so effectively the computing infrastructure essential to our analyses. Finally, we acknowledge the enduring support for the construction and operation of the LHC and the CMS detector provided by the following funding agencies: BMWFW and FWF (Austria); FNRS and FWO (Belgium); CNPq, CAPES, FAPERJ, and FAPESP (Brazil); MES (Bulgaria); CERN; CAS, MoST, and NSFC (China); COLCIENCIAS (Colombia); MSES and CSF (Croatia); RPF (Cyprus); SENESCYT (Ecuador); MoER, ERC IUT, and ERDF (Estonia); Academy of Finland, MEC, and HIP (Finland); CEA and CNRS/IN2P3 (France); BMBF, DFG, and HGF (Germany); GSRT (Greece); OTKA and NIH (Hungary); DAE and DST (India); IPM (Iran); SFI (Ireland); INFN (Italy); MSIP and NRF (Republic of Korea); LAS (Lithuania); MOE and UM (Malaysia); BUAP, CINVESTAV, CONACYT, LNS, SEP, and UASLP-FAI (Mexico); MBIE (New Zealand); PAEC (Pakistan); MSHE and NSC (Poland); FCT (Portugal); JINR (Dubna); MON, RosAtom, RAS, RFBR and RAEP (Russia); MESTD (Serbia); SEIDI, CPAN, PCTI and FEDER (Spain); Swiss Funding Agencies (Switzerland); MST (Taipei); ThEPCenter, IPST, STAR, and NSTDA (Thailand); TUBITAK and TAEK (Turkey); NASU and SFFR (Ukraine); STFC (United Kingdom); DOE and NSF (USA).
\end{acknowledgments}

\bibliography{auto_generated}
\ifthenelse{\boolean{cms@external}}{}{
\appendix
\section{Supplemental Material\label{app:suppMat}}

A proper comparison between the data presented in this Letter and theory can be done when the theory calculations smear the generator-level jet \pt to the resolution observed in detector-level jets. The procedure is given below.

The \ptGen is smeared into \ptReco by multiplying the \ptGen with a number sampled from a Gaussian distribution with mean $1$ and variance $\sigma^{2}(\ptGen)$. The functional form is the same for pp and PbPb :
\begin{equation*}
  \sigma\left(\ptGen\right) = \sqrt{C^{2}+{\frac{S^{2}}{\ptGen}}+{\frac{N^{2}}{\left( \ptGen \right)^2}}}
\end{equation*}

The parameter $S$ describes the \pt dependence of the jet energy resolution, 
$C$ represents the high-\pt limit of the resolution, and $N$ reflects the effect of UE fluctuations on the energy resolution. 
The parameters for $\sigma\left(\ptGen\right)$ are determined using {\PYTHIA{}+\HYDJET} (for \PbPb) and \PYTHIA (for $\Pp\Pp$) samples. The parameter $C$ is determined only from the \PYTHIA sample, and has the same value for $\Pp\Pp$ and \PbPb, $0.061\pm0.001$. The \PbPb ($\Pp\Pp$) numerical values for $S$ and $N$ are $1.24\pm0.04$ ($0.95\pm0.01$) and $8.08\pm0.15$ ($0.001\pm0.001$), respectively.

Arguably a more accurate comparison would include the smearing of generator-level jet $\phi$ as well. However, simulation studies showed that, compared to jet \pt smearing, jet $\phi$ smearing has negligible effect on the observables presented and therefore is omitted in this guide.

}

\cleardoublepage \section{The CMS Collaboration \label{app:collab}}\begin{sloppypar}\hyphenpenalty=5000\widowpenalty=500\clubpenalty=5000\textbf{Yerevan Physics Institute,  Yerevan,  Armenia}\\*[0pt]
A.M.~Sirunyan, A.~Tumasyan
\vskip\cmsinstskip
\textbf{Institut f\"{u}r Hochenergiephysik,  Wien,  Austria}\\*[0pt]
W.~Adam, E.~Asilar, T.~Bergauer, J.~Brandstetter, E.~Brondolin, M.~Dragicevic, J.~Er\"{o}, M.~Flechl, M.~Friedl, R.~Fr\"{u}hwirth\cmsAuthorMark{1}, V.M.~Ghete, C.~Hartl, N.~H\"{o}rmann, J.~Hrubec, M.~Jeitler\cmsAuthorMark{1}, A.~K\"{o}nig, I.~Kr\"{a}tschmer, D.~Liko, T.~Matsushita, I.~Mikulec, D.~Rabady, N.~Rad, B.~Rahbaran, H.~Rohringer, J.~Schieck\cmsAuthorMark{1}, J.~Strauss, W.~Waltenberger, C.-E.~Wulz\cmsAuthorMark{1}
% ...
\vskip\cmsinstskip
% ...
1:~~Also at Vienna University of Technology, Vienna, Austria\\
% ...

\end{sloppypar}
\end{document}